# Crypto-assets better safe-havens than Gold during Covid-19: The case of European indices

Alhonita YATIE[1]

**Abstract**

As the first crisis faced by Crypto-assets, Covid-19 updated the debate about their safe-haven properties. Our paper tries to analyze the safe-haven properties of Crypto-assets and Gold for European assets. We find that Gold has not been more efficient than Crypto-assets (Tether, Cardano and Dogecoin) as safe-haven during the market crash due to Covid-19 in March 2020. We also found that during the study period Bitcoin, Ethereum, Litecoin and Ripple were just diversifiers for the European indices. Finally, Tether, Cardano and Dogecoin showed hedging properties like Gold before and after the market crash.

[1] Bordeaux Sciences Economiques, alhonita.yatie@u-bordeaux.fr

## 1.1.Introduction

Since the World Health Organization's announcement on 11 March 2020 of a global pandemic due to Covid-19, the health crisis has turned into an economic and financial crisis causing several losses on many indices. DAX-30, , IMOEX, FTSE-100 and CAC40 have lost respectively -12.24%, -12.4%, -8.28%,-10.87% and -9%.This situation of a general decline in several financial indices is a reminder of the interconnections between financial markets. Given this interconnection it is necessary for an investor to select the assets that are negatively or weakly correlated to each other in order to reduce losses on his portfolio.

Traditionally, Gold has been use as safe-haven and several studies have shown that it is efficient in times of distress (Baur and Lucey, 2010 ; Reboredo, 2013; Baur and McDermott, 2010, 2016; Ji et *al.*, 2020 ). But during the crash caused by the pandemic, Gold prices also dropped (-3.51%). For Akhtaruzzaman et *al.* (2020), Gold was not a safe-haven during the market crash of March 2020 as its correlations with many assets were positive. Given the inability of Gold to perform its traditional function as a safe-haven asset during this crisis, we decide to study the properties of another asset class: Crypto-assets.

Regarding Crypto-assets, there is no consensus about their properties as safe-havens. Indeed, some authors consider Crypto-assets and Bitcoin in particular very useful in times of crisis because it is decoupled from the traditional assets (Bouri et *al.*, 2017; Gil-Alana et al., 2020) and exhibits clearly safe-haven properties against economic policy uncertainty or Covid-19 (Bouri et *al.*, 2019, Corbet et *al.*, 2020a; Goodell and Goutte, 2021a). Crypto-assets are also useful as a hedge against US dollar fluctuations, euro indices, equity market, or ETFs (Dyhrberg, 2016a; Bouri et *al.*, 2021; Chan et *al.*, 2019; Samah, 2020; Kristjanpoller et *al.*, 2020) and as diversification assets (Brière et *al.*, 2015; Platanakis and Urquhart, 2020) because their introduction in a portfolio would lower the overall risk, while other studies show that Crypto-assets are not safe-havens (Goodell and Goutte, 2021b (except Tether); Shahzad et *al.,* 2019; Colon and Gee, 2020) and can even be risky in a portfolio (Al-Khazali et *al.*, 2018; Corbet et *al.,* 2020b).

Based on this observation, we consider that Covid-19 crisis has created the right conditions to test the properties of Crypto-assets as safe-havens using a Dynamic Conditional Correlation (DCC-GARCH). As defined by Engle (2002), a DCC-GARCH has "*the flexibility of univariate GARCH but not the complexity of conventional multivariate GARCH*" and it also allows the correlations to change over the time. A DCC-GARCH is a practical tool for analyzing the evolution of the correlation between two financial assets.

We therefore, select some of the most famous Crypto-assets namely: Bitcoin (BTC), Ethereum (ETH), Cardano (ADA), Ripple (XRP), Dogecoin (DOGE), Tether (USDT) and Litecoin (LTC) ) and test their properties as safe-havens on several European indices: CAC40 (France), DAX-30 (Germany), IBEX-35 (Spain), FTSE-100 (England), FTSE-MIB (Italy), OMXS30 (Sweden), BIST-100 (Turkey), PSI-20 (Portugal), IMOEX (Russia) and BEL-20 (Belgium). To our knowledge, there is no study in the literature that thoroughly analyzes the European financial market during the market crash due to Covid-19 by comparing the safe-haven characteristics of Crypto-assets to those of Gold.

We find that Crypto-assets (namely Tether, Cardano and Dogecoin) were more efficient than Gold as safe-havens during the market crash (due to Covid-19), as their correlation with FTSE-100, DAX-30, , IMOEX, IBEX-35, BEL20, FTSE-MIB, IMOEX, PSI20 and OMXS30 was negative. Furthermore, Tether, Cardano and Dogecoin have shown hedging characteristics like Gold before and after the market crash. Finally, Bitcoin, Ethereum, Litecoin and Ripple were just diversifiers for the European indices.

The chapter is organized as follows: Section 2 presents the data and methodology, while section 3 describes the empirical results, and section 4 summarizes our findings and provides conclusions.

## 1.2.Materials and Methods

The data on Crypto-assets were collected in euros on yahoo finance, those on European indices were collected on investing.com, then the data on Gold were extracted on https://www.banque-france.fr/ . Our empirical analyzes are conducted with daily log returns covering the period from January $2^{nd}$ 2020 to June $30^{th}$ 2020. **Table 1** and **Table 2** present the descriptive statistics where we can notice that the highest variances are observe among the Crypto-assets: Cardano (6.86), Ethereum (6.22), Bitcoin (5.98), Litecoin (5.65), Ripple (4.81) and Dogecoin (3.52). The smallest variance goes to Tether (0.85).

**Table 1 : Descriptive statistics (Part 1)**

| | *BITCOIN (BTC)* | *CARDANO (ADA)* | *DOGECOIN (DOGE)* | *ETHEREUM (ETH)* | *LITECOIN (LTC)* | *TETHER (USDT)* | *RIPPLE (XRP)* | *GOLD* |
|---|---|---|---|---|---|---|---|---|
| *Mean* | -0.11 | 0.87 | 0.26 | 0.31 | 0.004 | -0.0003 | -0.05 | 0.12 |
| *Min* | -44.67 | -50.54 | -10.32 | -54.01 | -43.95 | -3.44 | -38.63 | -5.51 |
| *Max* | 19.35 | 28.61 | 15.1 | 20.15 | 19.51 | 5.71 | 17.02 | 5.11 |
| *Std. Dev.* | 5.98 | 6.86 | 3.52 | 6.22 | 5.65 | 0.85 | 4.81 | 1.34 |
| *Obs.* | 181 | 181 | 181 | 181 | 181 | 181 | 181 | 181 |

**Table 2 : Descriptive statistics (Part 2)**

| | *BEL20* | *BIST-100* | *CAC40* | *DAX-30* | *FTSE-100* | *FTSE MIB* | *IBEX-35* | *IMOEX* | *OMXS30* | *PSI 20* |
|---|---|---|---|---|---|---|---|---|---|---|
| *Mean* | -0.06 | 0.05 | -0.20 | -017 | -0.44 | -0.19 | -0.30 | -0.03 | -0.11 | -0.15 |
| *Min.* | -7.72 | -13.75 | -11.36 | -8.98 | -11.51 | -18.54 | -7.91 | -10.28 | -9.77 | -6.92 |
| *Max.* | 6.44 | 8.72 | 4.99 | 7.94 | 8.67 | 8.67 | 8.24 | 8.65 | 5.58 | 7.72 |
| *Std. Dev.* | 2.01 | 2.36 | 2.43 | 2.69 | 2.59 | 3.05 | 2.37 | 2.2 | 2.39 | 2.00 |
| *Obs.* | 181 | 181 | 181 | 181 | 181 | 181 | 181 | 181 | 181 | 181 |

Then we take a look into the pair-wise values between all the assets using a heat-map representation (**Figure 1**). A dark red color indicates that the respective two variables are highly negatively correlated, while dark blue indicates a highly positive correlation. For instance, we can see that Tether (USDT) is mostly negatively correlated with the other cryptocurrencies and indices. It can give a hint about the safe-haven property of Tether during this period. For the assets like Bitcoin (BTC), Ethereum (ETH), Litecoin (LTC), Ripple (XRP) and Gold where we observe a positive correlation with European indices, we can have a suspicious idea about their failure as safe-havens during this period.

Before using the DCC-GARCH, we test our variables for stationarity. Augmented Dickey-Fuller and Phillips-Perron tests indicate that all return series are stationary (**Table 3**). We also make a test about the presence of ARCH effects. ARCH test allows the rejection of the null hypothesis of an absence of Arch effects (**Table 4**).

Then we use the DCC-GARCH methodology developed by Engle (2002) to examine the dynamic correlation between Crypto-assets (and Gold) and financial market indices. The aim is to capture the dynamic nature of Gold and Crypto-assets as hedge, diversifier or safe-haven for European assets during the market crash of March 2020. DCC-GARCH captures the interactions among assets by allowing the correlations to change over the time. It is estimated in two steps – the first is a series of univariate GARCH estimates and the second the correlation estimate.

The model is defined as:

$$r_t = \mu_t + \varepsilon_t \quad , \varepsilon_t | \ E(\varepsilon_t) = 0, Cov\ (\varepsilon_t) = H_t \tag{1}$$

$$\varepsilon_t = \sqrt{H_t} u_t \ , \ u_t \sim N(0, I) \tag{2}$$

$$H_t = D_t R_t D_t \tag{3}$$

Where $r_t$, $\mu_t$, $\varepsilon_t$ and $u_t$ are $N \times 1$ dimensional vectors representing respectively log returns of $n$ assets at time $t$, expected value of the conditional $r_t$, mean-corrected returns of $n$ assets at time $t$ and $iid$ errors.

$H_t$, $R_t$ and $D_t$ are $N \times N$ dimensional matrices illustrating respectively time-varying matrix of conditional variances of $\varepsilon_t$, time-varying conditional correlation matrix of $\varepsilon_t$ and time-varying diagonal matrix of conditional standard deviations of $\varepsilon_t$.

At the first stage of the DCC, $D_t$ is generated by estimating a Bollerslev (1986) GARCH model for each series. We assume one shock and one persistency parameter:

$$h_{i,t} = \omega + \alpha_i \varepsilon_{i,t-1}^2 + \beta_i h_{i,t-1} \quad (4)$$

$h_{i,t}$ is the conditional variance of asset $i$ at time $t$ .

At the second stage, the dynamic conditional correlations are compute as follows:

$$R_t = diag(\sqrt{q_{iit}^{-1}}, \dots, \sqrt{q_{NNt}^{-1}})\ Q_t diag(\sqrt{q_{iit}^{-1}}, \dots, \sqrt{q_{NNt}^{-1}}) \quad (5)$$

$(\sqrt{q_{iit}^{-1}}, \dots, \sqrt{q_{NNt}^{-1}})$ is the square root of the diagonal elements of $Q_t$

The DCC-GARCH (1,1) equation is then given by $Q_t$:

$$Q_t = (1 - \alpha - \beta)\bar{Q} + a\varphi_{t-1}\varphi'_{t-1} + \beta Q_{t-1} \text{ with } \varphi_t = D_t^{-1}\varepsilon_t \quad (6)$$

Where $\varphi_t$ is a vector of standardized residuals from the first-step estimation of the GARCH (1,1) process, $Q_t$ is the time-varying unconditional correlation matrix of $\varphi_t$ and $\bar{Q}$ is a $N \times N$ dimensional positive-definite matrix which represents the unconditional covariance matrix of $\varphi_t$ .

$\alpha$ and $\beta$ satisfy $\alpha + \beta < 1$. As long as $\alpha + \beta < 1$ is fulfilled, $Q_t$ and $R_t$ are varying over the time.

As a robustness test, we run OLS regressions with Prais-Winstern robust estimator, as presented in equations 7 and 8 :

$$Coin_t = \gamma + \beta_1 Coin_{t-1} + \beta_2 Stock_t + \beta_3 Covid * Stock_t + \beta_4 Stock_{t-1} + \varepsilon_t \quad (7)$$

$$Gold_t = \gamma + \beta_1 Gold_{t-1} + \beta_2 Stock_t + \beta_3 Covid * Stock_t + \beta_4 Stock_{t-1} + \varepsilon_t \quad (8)$$

Where $Coin_t$ is the Crypto-asset return at day-t, $Gold_t$ is gold return at day-t, $Stock_t$ is stock return at day-t and $Covid * Stock_t$ is a dummy variable that equals one if day-t is on the pandemic announcement date (March 11, 2020) or the subsequent days (14 days). If the Crypto-asset or Gold serves as a safe-haven in the pandemic, then the coefficient of $\beta_3$ is expected to be negative (Baur et *al.*, 2018). *"A safe haven asset holds its value in 'stormy* *weather' or adverse market conditions. Such an asset offers investors the opportunity to protect wealth in the event of negative market conditions"* Baur and McDermott (2010).

**Figure1 : Heatmap of the correlation between the return series**

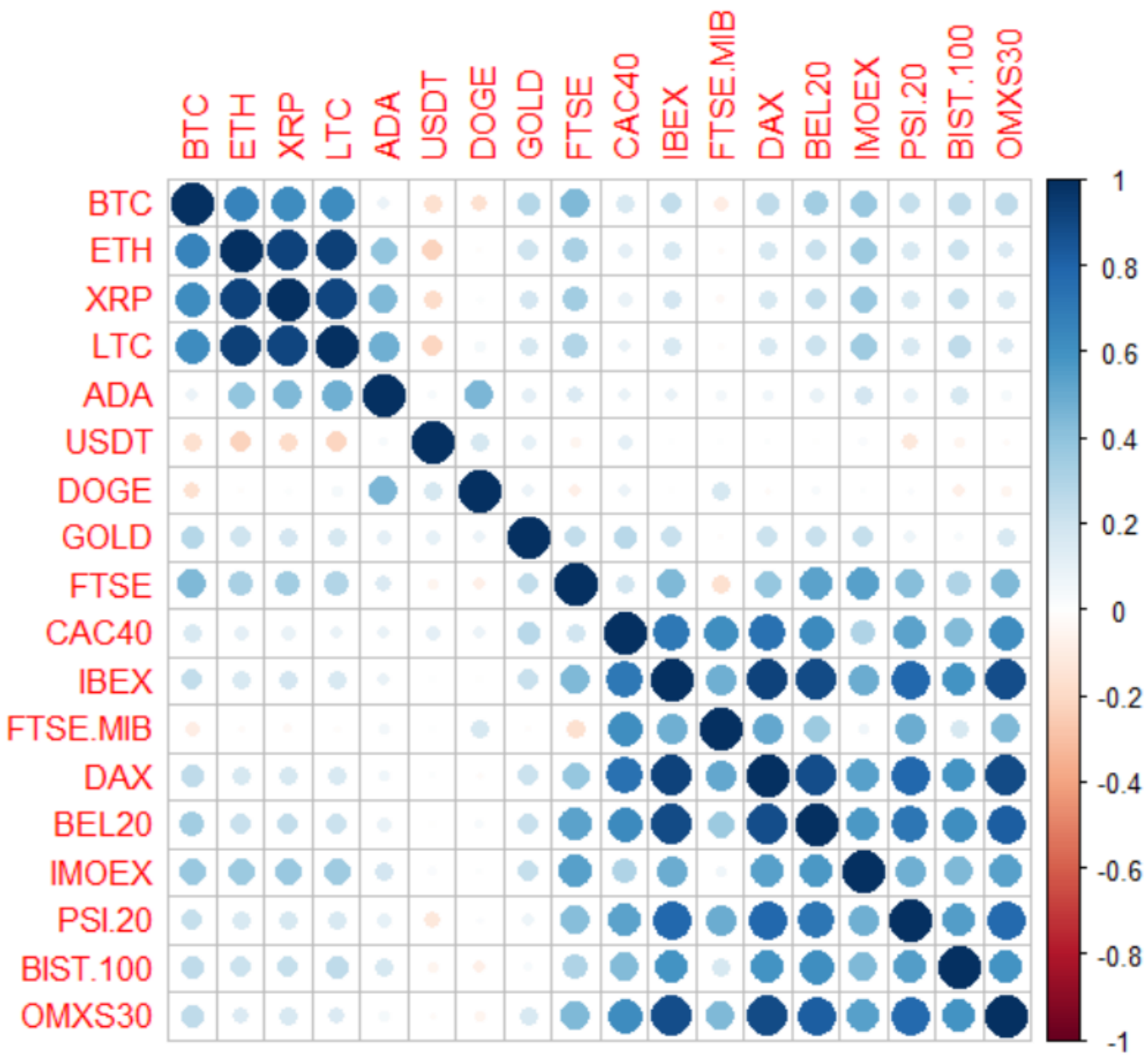

**Note: A dark red color indicates that the respective two variables are highly negatively correlate, while dark blue indicates a highly positive correlation.**

## 1.3.Results and Discussions

The European indices and Gold correlations (**Figure 2**) are increasing and positive after the market crash due to the WHO's announcement of March $11^{th}$ 2020 (symbolized by the red line). The correlations remained positive during few days after the market crash and Gold returned to its traditional function of hedge later. We can clearly observe that Gold has failed in its traditional function as a safe-haven asset right after the WHO's announcement and the subsequent days (almost 10 days, except for CAC 40, BIST-100 and FTSE MIB)). During the market crash, the correlation between Gold and European indices immediately rose, reducing Gold to the status of diversifier. The failure of Gold as a safe-haven is confirmed by our OLS regression (**Table 5**). The coefficient about COVID-19 crisis is positive and significant for all the indices (except FTSE-100 and FTSE MIB). Even if Gold showed in **Figure 2** characteristics of a safe-haven asset for some indices, such as CAC 40, BIST-100 and FTSE MIB, its reaction is much too brief to be considered significant by the OLS model.

These results are in line with Akhtaruzzaman et *al.* (2020) who use a DCC-GARCH to exhibit that Gold was not a safe-haven during the period from March $17^{th}$ to April $24^{th}$ 2020 as the correlations between Gold and the selecte assets (S&P 500, NIKKEI 225 etc.) were positive.

**Figure2: Correlations between Gold and European indices**

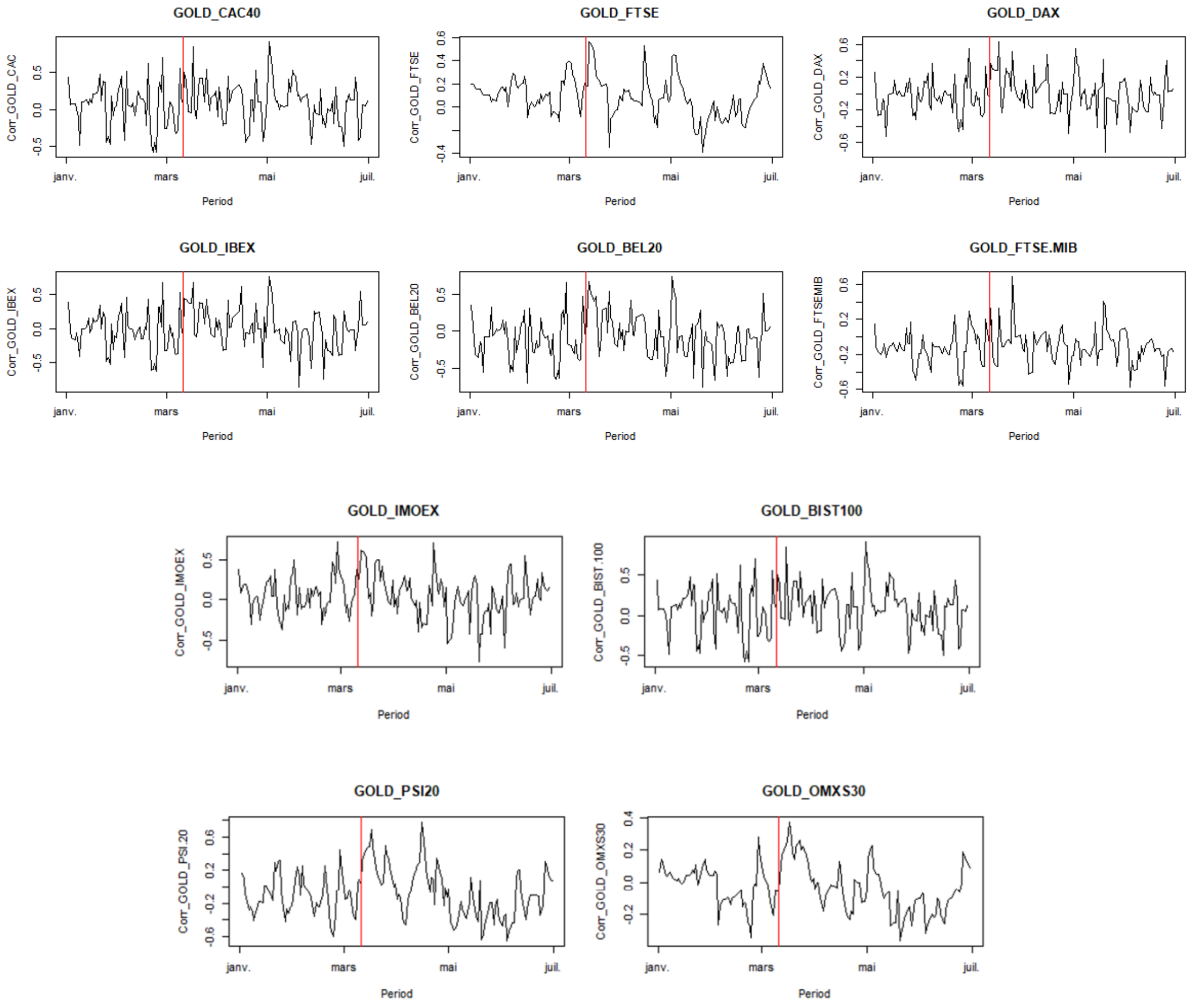

**Figure3: Correlations between Bitcoin and European indices**

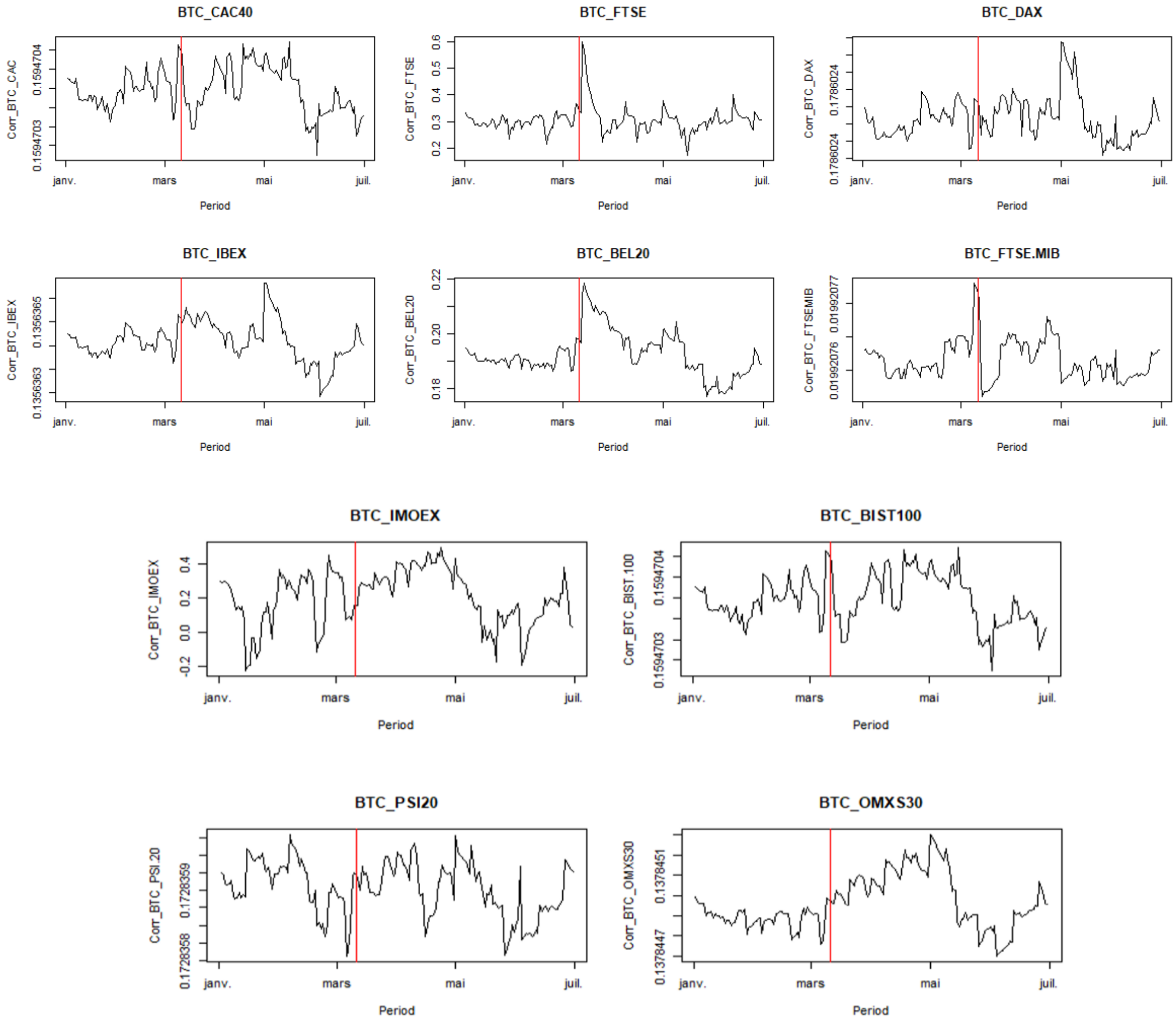

When we take a look to Crypto-assets, Bitcoin (**Figure 3**) did not exhibit safe-haven properties during the market crash. After the WHO's announcement most of the couple correlation increased and all were positive. During the period of study we can notice that Bitcoin was just a diversifier except for IMOEX where it showed sometimes hedging features. These results are in contradiction with a lot of studies in the literature (Corbet et *al.*, 2018; Shahzad et *al.*, 2019; Kumar, 2020; Dwita Mariana et *al.*, 2021) where Bitcoin is depicted as a safe-haven because of its decoupling of the traditional financial market or its weak correlation with the traditional assets. According to Corbet et *al.* (2018) Crypto-assets namely Bitcoin, Ripple and Litecoin are isolated and disconnected from mainstream assets. Therefore, these crypto-assets would be useful to diversify or hedge a portfolio composed with mainstream assets. As we saw with the correlation heat-map, the correlation between Bitcoin and the European indices is in most cases positive and it will be more important with the growing usage of Bitcoin. Bitcoin cannot act as safe-haven or hedge for the European indices because of this positive relationship. To a lesser extent, Bitcoin can be a diversifier for the European indices. **Table 5** confirmed the absence of a safe-haven property for Bitcoin during the WHO's announcement and 14 days later. The OLS results show globally a positive relationship between Bitcoin returns and the Covid-19 variable.

**Figure 4: Correlations between Ethereum and European indices**

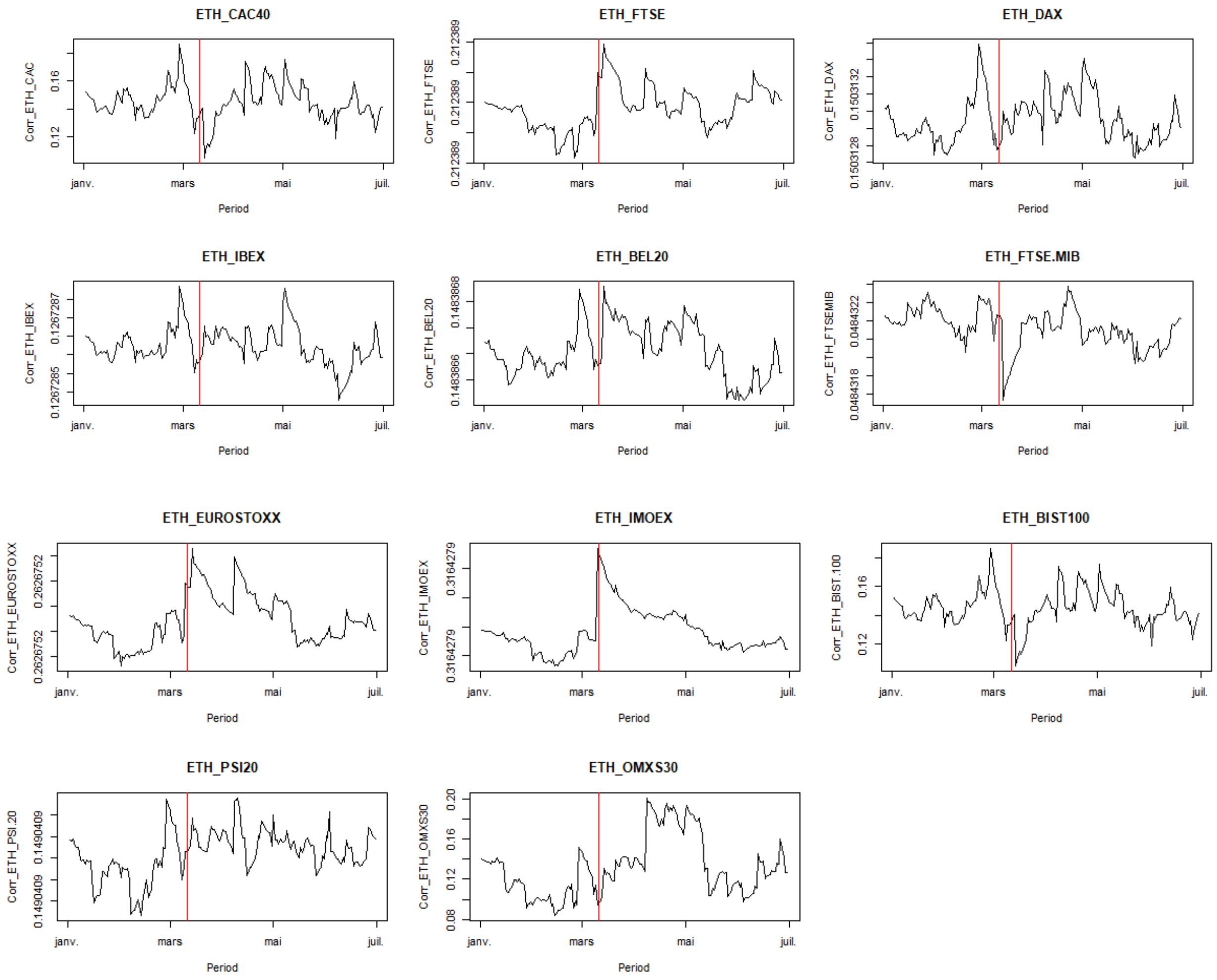

**Figure 5: Correlations between Litecoin and European indices**

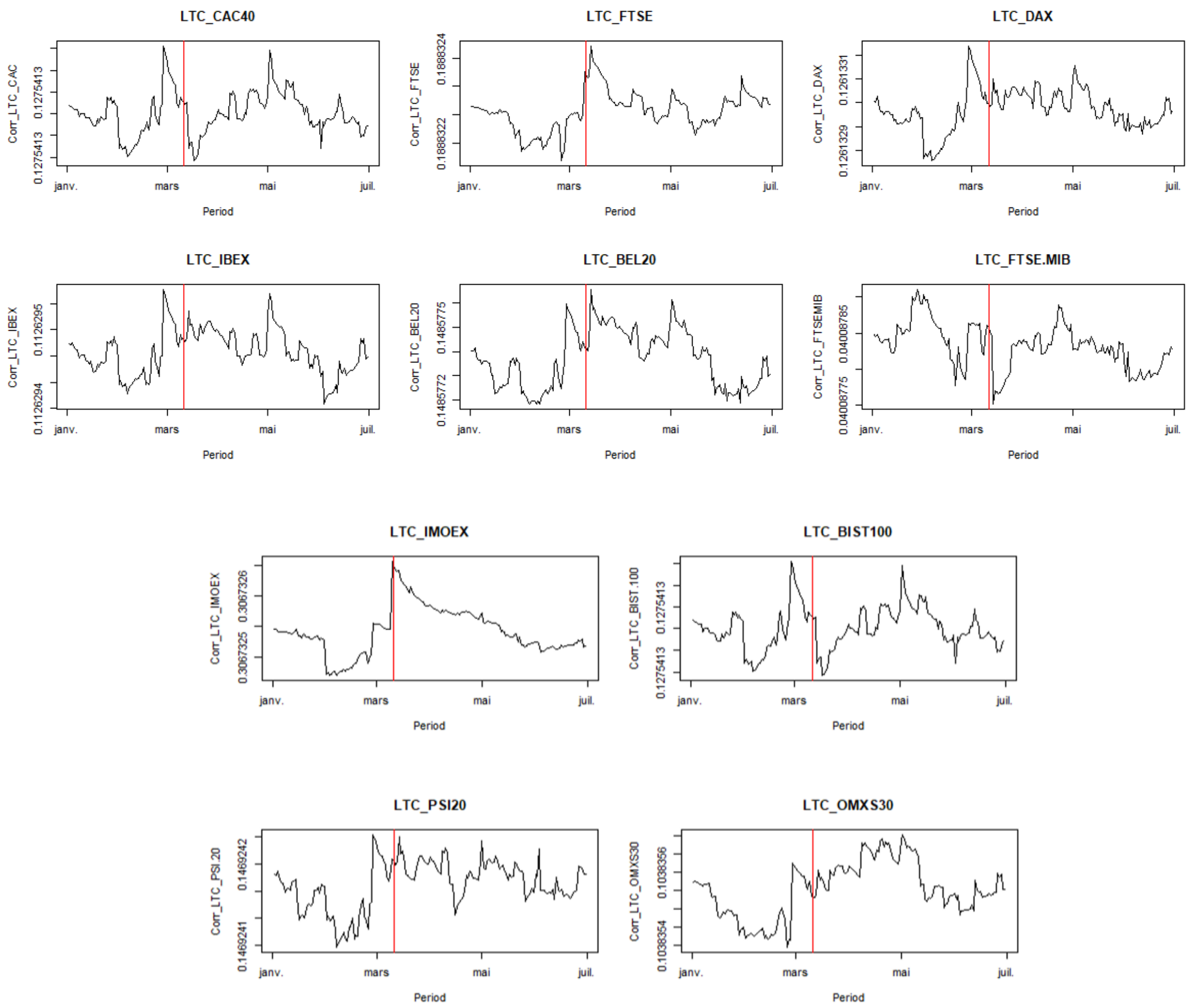

We find the same results for Ethereum (**Figure 4**), Litecoin (**Figure 5**) and Ripple (**Figure 6**). None of these Crypto-assets acted as a safe-haven as all the correlations provided by the DCC-GARCH were positive. We also notice the positive relationship between these Crypto-assets and the European indices on the heat-map (**Figure 2**). The OLS results corroborate the absence of safe-haven characteristics for Ethereum, Litecoin and Ripple as their relationship with the Covid-19 dummy variable is positive and significative in some cases (**Table 5**). These results are in line with Colon and McGee (2020), Goodell and Goutte (2021b) and Corbet et *al.* (2020b). Ethereum, Litecoin and Ripple were also weak diversifiers according for the European indices because of the positive and stable correlation between them.

**Figure 6: Correlations between Ripple and European indices**

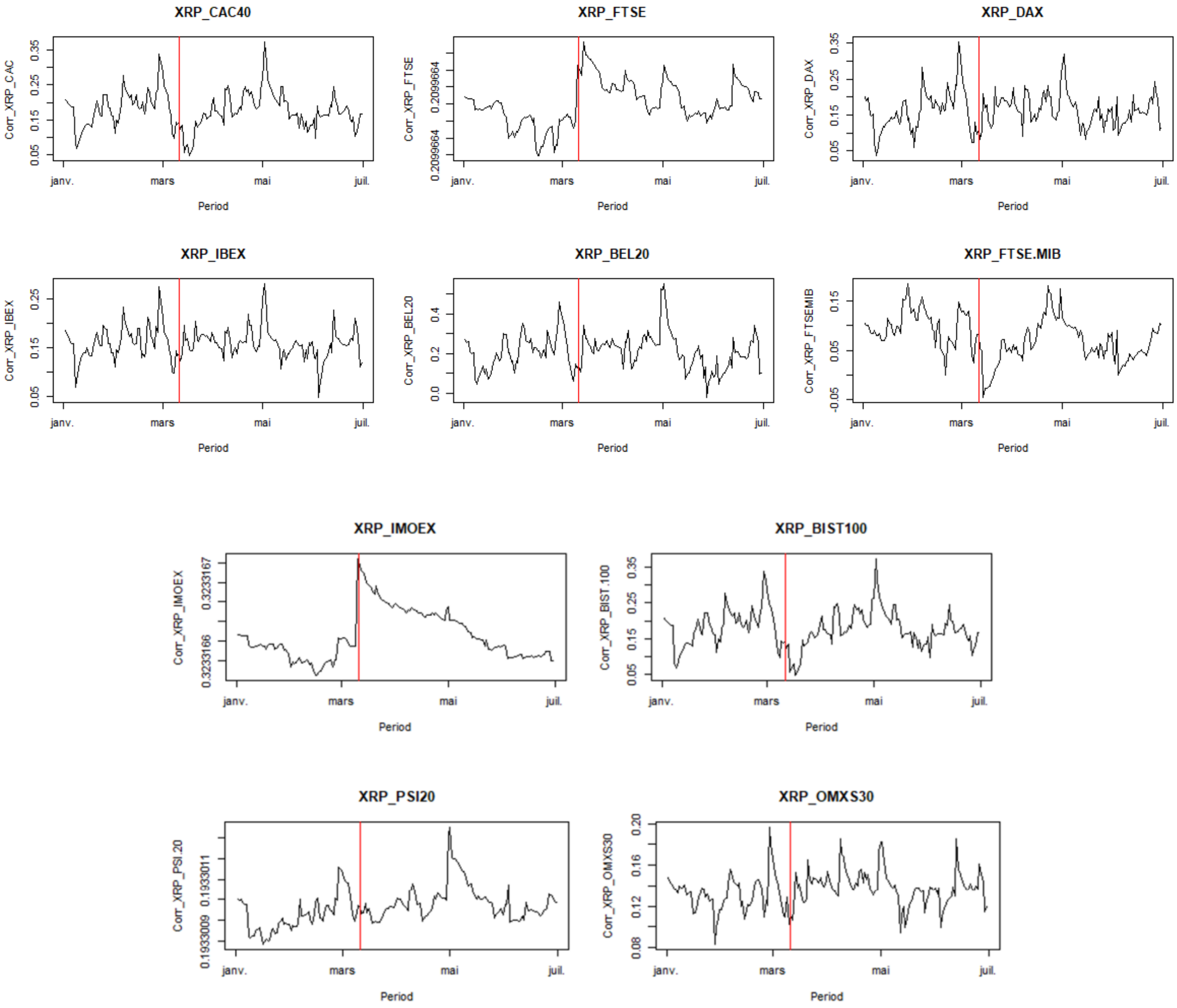

**Figure 7** shows the correlations between Tether and European indices. As we can see, Tether acted as a safe-haven for all the indices (except CAC40) because the correlation decreased and was negative after the WHO's announcement and the beginning of the market crash.

Before and after the market crash, the correlations were globally negative (except for CAC40). Tether shows some hedging features like Gold for European indices during the period of study. These results are in line with Goodell and Goutte (2021b). They explain the safe-haven property of Tether by its stability with dollar. The results of the DCC-GARCH are also supported by the OLS regression (**Table 5**). The relationships between Tether returns and the Covid-19 dummy variable are negative, confirming the safe-haven property of Tether.

**Figure 7: Correlations between Tether and European indices**

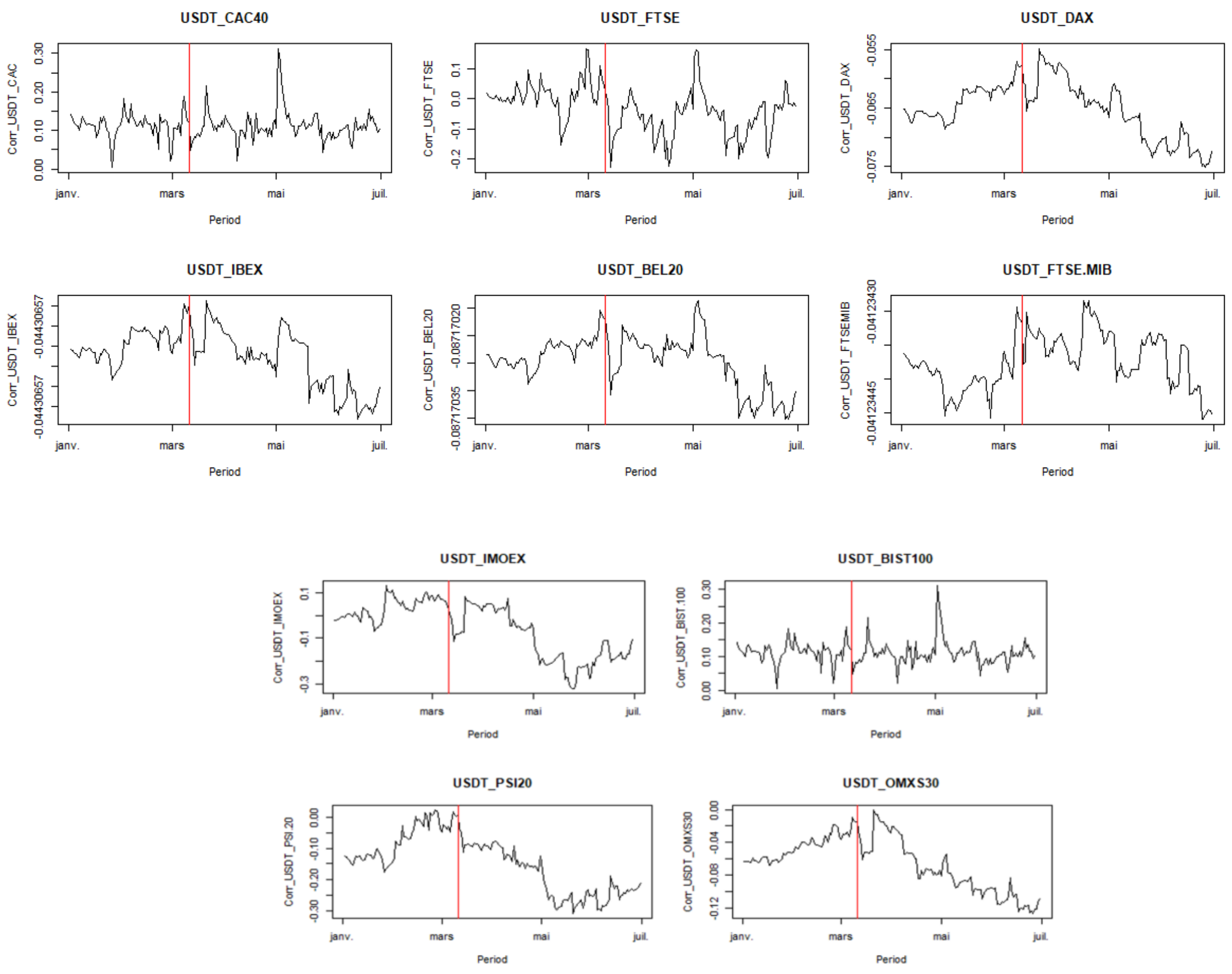

**Figure 8** shows the correlations between Cardano and European indices. Cardano was a safe-haven for all the indices except for CAC40, IMOEX, IBEX-35, and BIST-100. Unlike Tether, the correlation decreasing was not always automatic (e.g. FTSE 100) and drastic. The drops were small in magnitude compared to those of Tether. However, Cardano was also a hedge for most of the European indices. The results of the OLS regression are also mixed (**Table 5**) confirming that Cardano was not a safe-haven for CAC 40, IMOEX, IBEX-35, and BIST-100.

**Figure 8: Correlations between Cardano and European indices**

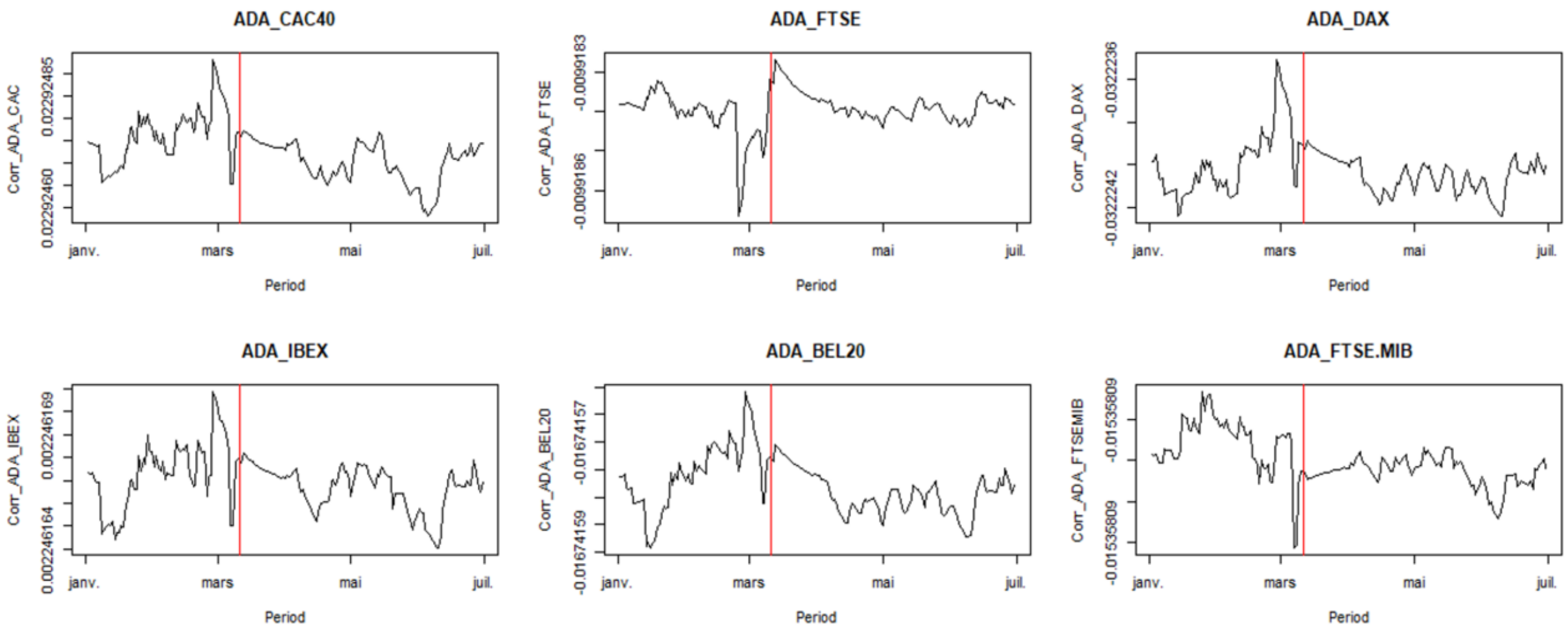

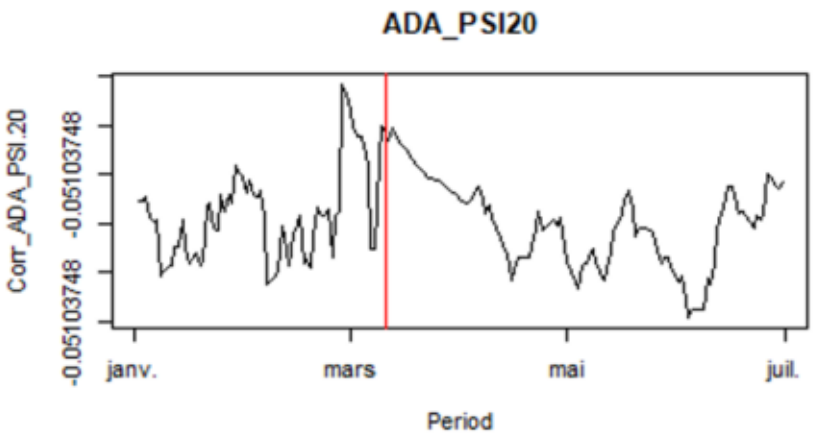

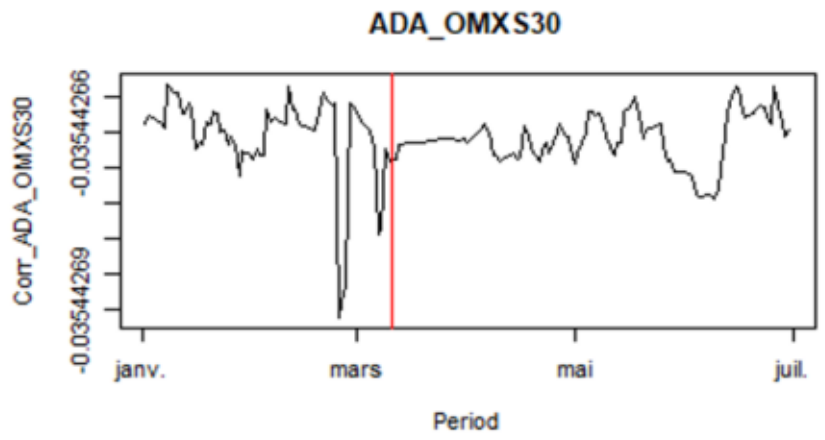

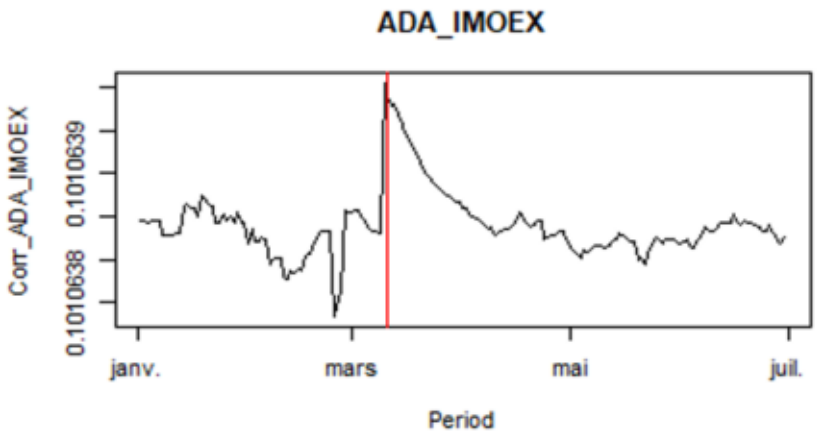

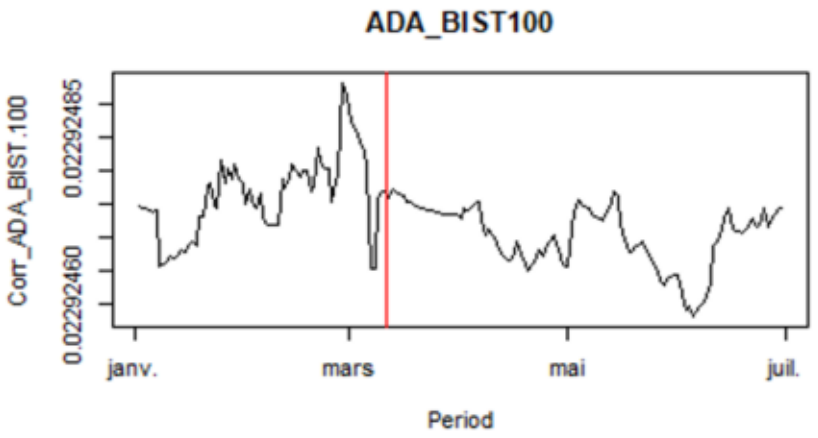

About Dogecoin (**Figure 9**), it was a safe-haven for OMSX30 where the correlation were negative during the market crash. For FTSE-100, DAX and , the correlations dropped but became negative few days after the WHO's announcement. Like Tether and Cardano, Dogecoin showed hedging  properties for these indices. But, Tether appeared as a better safe-haven than Cardano and Dogecoin during this period of study. Finally, our OLS regression shows that Dogecoin has been a safe-haven only for  OMXS30 and FTSE-100 (**Table 5**). For DAX the coefficient is negative but not significant.

**Figure 9: Correlations between Dogecoin and European indices**

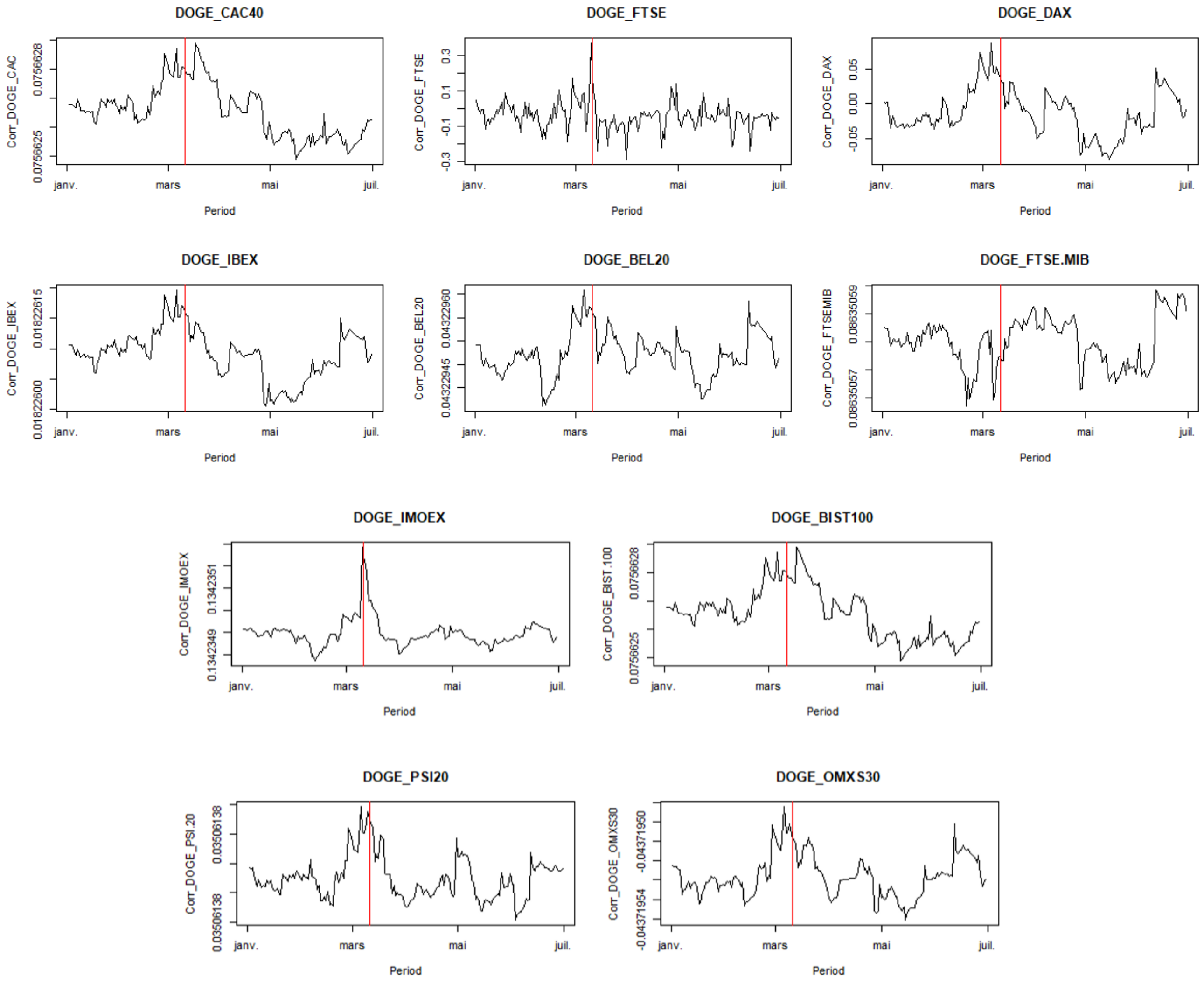

As one of the reasons why Cardano and Dogecoin were effective as safe-havens during the Covid-19 crisis, we identified their market capitalization. As example the market capitalizations of Cardano and Dogecoin were respectively (around) 780 million and 220 million versus 100 billion for Bitcoin and 25 billion for Ethereum in March 2020. As their usage will grow, they will be more implemented in the traditional financial market and we think that they will probably lose as Bitcoin or Ethereum their safe-haven ( or hedging ) properties.

**Table3: Augmented-Dickey Fuller results**

| *VARIABLES* | *T-STATISTIC* | *CONCLUSION* |
|---|---|---|
| ***BITCOIN*** | -10.909*** | Stationary |
| ***CARDANO*** | -16.673*** | Stationary |
| ***DOGECOIN*** | -13.989*** | Stationary |
| ***ETHEREUM*** | -5.396*** | Stationary |
| ***LITECOIN*** | -16.395*** | Stationary |
| ***RIPPLE*** | -5.760*** | Stationary |
| ***TETHER*** | -16.614*** | Stationary |
| ***GOLD*** | -8.705*** | Stationary |
| | | |
| ***BEL-20*** | -7.678*** | Stationary |
| ***BIST-100*** | -10.228*** | Stationary |
| ***CAC 40*** | -8.621*** | Stationary |
| ***DAX-30*** | -9.668*** | Stationary |
| | -5.067*** | Stationary |
| ***FTSE-100*** | -8.684*** | Stationary |
| ***FTSE MIB*** | -10.362*** | Stationary |
| ***IBEX-35*** | -9.142*** | Stationary |
| ***IMOEX*** | -10.029*** | Stationary |
| ***OMXS30*** | -10.185*** | Stationary |
| ***PSI 20*** | -9.277*** | Stationary |

**Note: The null hypothesis is the variable has a unit root (are not stationary). *** denotes rejection of the null hypothesis at the 0.01 level. All the results are confirmed by the Phillips-Perron test.**

## Table 4: ARCH-LM test and Heteroskedasticity test

| *VARIABLES* | *ARCH-LM* | *HETEROSKEASTICITY TEST* |
|---|---|---|
| ***BITCOIN*** | 4.371*** | 66.467*** |
| ***CARDANO*** | 9.2*** | 3.358** |
| ***DOGECOIN*** | 0.296** | 2.547*** |
| ***ETHEREUM*** | 8.443*** | 21.479*** |
| ***LITECOIN*** | 7.181*** | 6.408*** |
| ***RIPPLE*** | 8.402*** | 5.862*** |
| ***TETHER*** | 9.387*** | 5.307*** |
| ***GOLD*** | 6.292*** | 2.213** |

**Note: : For the ARCH LM test, null hypothesis is the absence of volatility clustering. The heteroskedasticity test is a Breusch-Pagan Godfrey test with a null hypothesis of homoscedasticity. *** significant at 1% level, ** significant at 5% level, * significant at 10% level.**

## Table5: OLS regression results

| | *BITCOIN* | *CARDANO* | *DOGECOIN* | *ETHEREUM* | *LITECOIN* | *RIPPLE* | *TETHER* | *GOLD* |
|---|---|---|---|---|---|---|---|---|
| *COVID*BEL-20* | 1.16* | 0.65 | -0.16 | 0.94* | 0.66* | 0.47 | 0.02 | 0.27*** |
| *COVID*BIST-100* | 1.82** | 0.46 | -0.38 | 1.42** | 0.99* | 1.01** | 0.02 | 0.34** |
| *COVID*CAC 40* | 0.32 | 0.53 | -0.07 | 0.14 | 0.04 | -0.21 | 0.01 | 0.2*** |
| *COVID*DAX-30* | 0.58** | 0.1 | -0.18 | 0.56 | 0.46 | 0.23 | 0.04 | 0.28*** |
| *COVID*FTSE-100* | 1.11* | -0.039 | -0.65** | 0.29 | 0.16 | 0.26 | -0.02 | -0.004 |
| *COVID*FTSE MIB* | -0.43 | -0.33 | 0.07 | -0.44 | -0.4 | -0.61 | 0.005 | -0.12 |
| *COVID*IBEX-35* | 1.01** | 0.22 | -0.29 | 0.93** | 0.71* | 0.45 | -0.008 | 0.3** |
| *COVID*IMOEX* | 1.21* | -0.17* | 0.03 | 0.12** | -0.07 | 0.08* | 0.016 | 0.01** |
| *COVID*OMXS30* | 0.8* | 0.2 | -0.16*** | 0.76 | 0.62 | 0.44 | 0.02 | 0.28*** |
| *COVID* PSI 20* | 0.98 | 0.4 | -0.48 | 0.57 | 0.35 | 0.19 | -0.02 | 0.48*** |
| | | | | | | | | |
| *BEL-20* | 0.15 | 0.84* | 0.16 | 0.05 | 0.15 | 0.23 | -0.01 | -0.06 |
| *BIST-100* | 0.11 | 0.1 | 0.01 | 0.23 | 0.36* | 0.27* | -0.03 | -0.08 |
| *CAC 40* | 0.07 | 0.18 | 0.2* | 0.07 | 0.09 | 0.16 | 0.07** | 0.06 |
| *DAX-30* | 0.18 | 0.19 | 0.07 | 0.15 | 0.15 | 0.23 | -0.002 | -0.04 |
| *FTSE-100* | 0.48*** | 0.53* | 0.18 | 0.71*** | 0.65*** | 0.56*** | 0.01 | 0.04 |
| *FTSE MIB* | -0.13 | 0.27 | 0.17 | 0.1 | 0.11 | 0.18 | 0.006 | -0.03 |
| *IBEX-35* | 0.06 | 0.17 | 0.1 | 0.08 | 0.15 | 0.24 | 0.02 | -0.004 |
| *IMOEX* | 0.41*** | 0.38 | 0.07 | 1.20*** | 0.98*** | 0.88*** | 0.03 | 0.06* |
| *OMXS30* | 0.12 | 0.17 | -0.01 | 0.07 | 0.1 | 0.15 | -0.01 | -0.04 |
| *PSI 20* | 0.77 | 0.26 | 0.24** | 0.25 | 0.31 | 0.36* | -0.03 | -0.14 |
| | | | | | | | | |
| *BEL-20 (-1)* | 0.23 | 0.19 | -0.1 | 0.39 | 0.23 | 0.23 | 0.01 | 0.05 |
| *BIST-100 (-1)* | -0.053 | 0.09 | -0.06 | 0.1* | 0.03 | 0.04 | 0.001 | -0.001 |
| *CAC 40( -1)* | 0.51 | 0.38 | -0.1 | 0.48 | 0.38 | 0.34 | -0.05* | -0.02 |
| *DAX-30 (-1)* | 0.23 | 0.34 | -0.07 | 0.23 | 0.16 | 0.13 | -0.02 | 0.01 |
| *FTSE-100 (-1)* | -0.28* | 0.22 | 0.25** | 0.09 | 0.005 | 0.03 | -0.05 | 0.05* |
| *FTSE MIB (-1)* | 0.66** | 0.6 | 0.03 | 0.13* | 0.16 | 0.08 | 0.01 | -0.005 |
| *IBEX-35 (-1)* | 0.37 | 0.51 | -0.01 | 0.3 | 0.17 | 0.12 | -0.02 | -0.003 |
| *IMOEX (-1)* | -0.18 | 0.04** | -0.06 | -0.02 | -0.03 | -0.11 | -0.01* | 0.03 |
| *OMXS30 (-1)* | 0.41 | 0.58 | 0.017 | 0.28 | 0.2 | 0.18 | -0.002 | 0.01 |
| *PSI 20 (-1)* | 0.5 | 0.67 | -0.02 | 0.36 | 0.27 | 0.19 | -0.04 | 0.01 |

**Note: Regression (OLS with Prais-Winstern robust estimator) results analyzing Crypto-assets and Gold as safe-havens based on Equations 7 and 8. and $Covid * Stock_t$ is a dummy variable that equals one if day-t is on the pandemic announcement date (March 11, 2020) or the subsequent days (14 days). days. If the Crypto-asset (or Gold) serves as a safe-haven, then the coefficient is expected to be negative and significant. *** significant at 1% level, ** significant at 5% level, * significant at 10% level.**

## 1.4.Conclusion

Covid-19 was the first crisis experienced by Crypto-assets. This crisis made it possible to concretely analyze their characteristics as safe-havens, so we were thus able to study their relationships with European indices via a DCC-GARCH methodology. We notice that Crypto-assets (Tether, Cardano and Dogecoin) were more efficient than Gold during this financial crisis. They exhibited interesting safe-haven characteristics for FTSE-100, DAX-30, IBEX-35, BEL20, FTSE-MIB, IMOEX, PSI20 and OMXS30. Moreover, before and after the market crash, Tether, Cardano, and Dogecoin even showed hedging properties like Gold.

We find no safe-haven properties for Bitcoin, Ethereum, Ripple and Litecoin. Those assets were just diversifiers during this crisis. These results are in line with Colon and McGee (2020), Goodell and Goutte (2021b) and Corbet et *al.* (2020b). This empirical study may provide investors with valuable information to make the best decisions about their portfolio allocation and can be interesting for academics.